\documentclass[prd,floatfix,preprintnumbers]{revtex4}
\usepackage{amssymb}
\usepackage{amsmath}
\usepackage{graphicx}
\usepackage{dcolumn}

\newcommand {\ga} {\ {\raise-.5ex\hbox{$\buildrel>\over\sim$}}\ }
\newcommand {\la} {\ {\raise-.5ex\hbox{$\buildrel<\over\sim$}}\ }

\begin{document}
\title{What do we learn by mapping dark energy to a single value of $w$?}
\author{Samuel S. Taylor}
\affiliation{Department of Physics and Astronomy, Vanderbilt
  University, Nashville, TN, 37235, USA}
\author{Robert J. Scherrer}

\affiliation{Department of Physics and Astronomy, Vanderbilt
  University, Nashville, TN, 37235, USA}
	
\begin{abstract}
We examine several dark energy models with a time-varying equation of state parameter, $w(z)$, to determine what
information can be derived by fitting the distance modulus in such models to a constant equation of state
parameter, $w_*$.  We derive $w_*$ as a function of the model parameters for the
Chevallier-Polarski-Linder (CPL) parametrization, and for the Dutta-Scherrer approximation to hilltop quintessence
models.  We find that all of the models examined here can be well-described by a pivot-like redshift,
$z_{pivot}$ at which the value of $w(z)$ in the model is equal to $w_*$. However, the exact value of $z_{pivot}$
is a model-dependent quantity; it varies from $z_{pivot} = 0.22-0.25$ for the CPL models to $z_{pivot} = 0.17-0.20$
for the hilltop quintessence models.
Hence, for all of the models considered here, a constant-$w$ fit gives the value of $w$ for $z$ near 0.2.  However,
given the fairly wide variation in $z_{pivot}$ over even this restricted set of models, the information gained by
fitting to a constant value of $w$ seems rather limited.
\end{abstract}
	
\maketitle
	
\section{Introduction}

Observational data \cite{union08,hicken,Amanullah,Union2,Hinshaw,Ade,Betoule} indicate that the
energy density in the universe consists of approximately 70\% dark energy, with negative pressure, 
and roughly 30\% pressureless matter, which includes both baryons
and dark matter.
The dark energy component can be characterized by its equation of state parameter $w$, which
is defined as the ratio of the dark energy pressure to its density:
\begin{equation}
\label{w}
w=p/\rho.
\end{equation}
The special case $w=-1$ and $\rho = constant$ corresponds to a cosmological constant, 
$\Lambda$.
Although current observations are consistent with a cosmological constant,
a dynamical equation of state is not ruled out.  One set of alternatives to $\Lambda$ consists of 
quintessence models, in which the dark energy originates from a scalar field $\phi$ with an associated potential $V(\phi)$
\cite{RatraPeebles,Wetterich,Ferreira,CLW,CaldwellDaveSteinhardt,Liddle,SteinhardtWangZlatev}.
(See Ref. \cite{Copeland1} for a review).  Generically in such models, $w$ evolves with time.
It has long been known that even perfect information about supernova luminosity distances
is insufficient to determine the exact evolution of $w$ \cite{Maor1,Wolf1}, but most studies assume
a restricted form for the evolution of $w$ as a function of the redshift $z$ (or, equivalently,
the scale factor $a$).

In fitting observational data to dark energy models beyond $\Lambda$CDM, the first step is often to fit the data
to a fixed single value for $w$.  While this approach has the virtue of simplicity, constant-$w$ models, aside from $\Lambda$CDM,
are not necessarily the most well-motivated.  However, given the ubiquity of this approach, it is reasonable to try to understand the information that
a constant-$w$ fit provides when it is applied to dark energy models in which $w$ is not constant but instead varies with time.

In moving beyond a fit to constant $w$, the most common approach is to model the behavior of $w(a)$ through the Chevallier-Polarski-Linder (CPL) parametrization,
in which $w$ is taken to be a linear function of the scale factor, namely \cite{CP,Linder}
\begin{equation}
\label{CPL}
w = w_0 + w_a (1-a),
\end{equation}
where $w_0$ and $w_a$ are constants.  This approach has several advantages. The CPL expression for $w$ is well-behaved
over the entire range from $a=0$, at which $w = w_0 + w_a$, all of the way to the present ($a=1$), at which
$w = w_0$.  Furthermore, Eq. (\ref{CPL}) can provide a good fit to the evolution of $w$ in a variety of scalar field models.
However, some limitations to this approach have
also been noted \cite{Wolf1,ScherrerCPL,Shlivko,Wolf2,Gialamas}.

Our study is similar in spirit to that of Shlivko and Steinhardt \cite{Shlivko}, who examined the ``best-fit" CPL models for a variety of quintessence
models and mapped those models onto the $w_a - w_0$ plane.  However, we are examining here a very different question:  how do
quintessence models with time-varying $w$ map onto a constant value of $w$?
Given an infinite number of time-varying $w$ models, one cannot, of course, investigate all of them.
An obvious first choice are the models described by the CPL parametrization.  For a second set of models, with nonlinear variation
in $w$, we have chosen
to examine the thawing hilltop quintessence models presented in Ref. \cite{ds1}.  These models have several
advantages for a study of this kind:
they represent a physically-motivated set of quintessence models, and they are characterized by a small number of free parameters,
but even this limited set of models diverges from both constant-$w$ and linear (CPL) evolution.  Thus, these models form a convenient test bed
for our study.

A previous discussion of the problems produced by fitting to a constant $w$ model was given by Maor et al. \cite{Maor2}, who assumed
an underlying fiducial model in which $w$ varied linearly with $z$.  They fit this fiducial model to a variety of assumed models for $w$, including
the case of constant $w$.  Their work highlighted the errors produced when $w$ is taken to be constant.

In the next section we examine these two sets of models with time-varying $w$.  Assuming that these models with time-varying $w$ represent the ``true" evolution of dark energy in the universe, we then determine, for
each model, the constant $w$
dark energy model that best fits the corresponding luminosity distances.  Our results are discussed in Sec. III.  We find that, in general,
a fit to constant $w$ simply gives the value of the dark energy equation of state at a pivot-like redshift of $z_{pivot} \approx 0.2$, but there is fairly
wide variation in the value of this redshift over the models we examine. 

\section{Methods and Results}

Any evolving dark energy model can be mapped onto a best fit constant value for $w$.  In this investigation, we use
the standard expression for the distance modulus $\mu$ as a function of redshift $z$,
\begin{equation}
\mu(z) = 5 \log_{10} D_L(z) + 42.38 - 5 \log_{10} h,
\end{equation}
where $h$ is the Hubble parameter in units of 100 km sec$^{-1}$ Mpc$^{-1}$ (which will drop out of our analysis)
and $D_L(z)$
is the dimensionless luminosity distance, given by
\begin{equation}
D_L(z) = (1+z) \int_0^{z} \frac{H_0}{H(z^\prime)} dz^\prime.
\end{equation}
Here $H(z^\prime)$ is the Hubble parameter at redshift $z^\prime$ and $H_0$ is its present-day value.
For a model containing nonrelativistic matter and a dark energy component with a constant equation of state parameter $w_*$,
the luminosity distance can be expressed as
\begin{equation}
\label{DL}
D_L(z) = (1+z) \int_0^{z} \left[\Omega_{M0} (1+ z^\prime)^3 + \Omega_{\phi 0} (1+
z^\prime)^{3(1+w_*)}\right]^{-1/2} dz^\prime,
\end{equation}
where $\Omega_{M0}$ and $\Omega_{\phi 0}$ are the present-day densities of matter and dark energy
normalized to the critical density.
If we have instead a time-varying equation of state, we get
\begin{equation}
\label{tildeDL}
\widetilde D_L(z) = (1+z) \int_0^{z} \left[\Omega_{M0} (1+ z^\prime)^3
+ \Omega_{\phi 0} \rho_\phi(z^\prime)/\rho_{\phi0}\right]^{-1/2} dz^\prime,
\end{equation}	
where $\rho_\phi/\rho_{\phi 0}$ is given in terms of $w$ by
\begin{equation}
\label{rho(z)}
\rho_\phi(a)/\rho_{\phi0} = \exp\left\{3\int_{a^{\prime} =a}^{1} [1+w(a^{\prime})] \frac{da^{\prime}}
{a^{\prime}}\right\},
\end{equation}
with $a = 1/(1+z)$.  In what follows we will assume throughout that $\Omega_{M0} + \Omega_{\phi 0} = 1$ with $\Omega_{M0} = 0.3$
and $\Omega_{\phi 0} = 0.7$.
Given any prescription for $w$ as a function of $z$, our procedure is to calculate $\widetilde D_L(z)$ using Eqs. (\ref{tildeDL})
and (\ref{rho(z)}) and then determine the value of $w_*$ in Eq. (\ref{DL}) that minimizes
the quantity $\chi^2(w_*)$ given by
\begin{equation}
\chi^2(w_*) = \int_{z=0}^{z_{max}} \left[ \log_{10} D_L(z) -  \log_{10} \widetilde D_L(z)\right]^2 dz.
\end{equation}
Roughly speaking, this is equivalent to performing a least-squares best fit for $\mu(z)$ assuming
a constant value of $w$
($w_*$ in Eq. \ref{DL}) when the
the ``true" $\mu(z)$ is given by Eqs. (\ref{tildeDL})
and (\ref{rho(z)}), in the limit where we have perfect information for the distance modulus between $z=0$ and $z=z_{max}$.
We take $z_{max} = 2$ throughout.  This is a somewhat arbitary choice, but we
do not expect our final results to be very sensitive to $z_{max}$, since for all of the models examined here, the
dark energy component is subdominant at high redshift.

Our procedure differs in several respects from the fitting procedure of Shlivko and Steinhardt \cite{Shlivko} in their
study of the CPL parametrization.  They fitted
$H(z)$ rather than $\mu(z)$, and their best fit criterion was the minimization of the maximum difference between
the true $H(z)$ (assumed to be a quintessence model) and the $H(z)$ given by the CPL parametrization over the redshift
range $z < 4$.  However, these particular choices are unlikely to yield significantly different results from our own approach.
The major difference, of course, is that Ref. \cite{Shlivko} is an investigation of the best-fit CPL parametrization,
while we are interested in the best-fit constant-$w$ model.

We now need a representative set of dark energy models with time-varying $w$ in order to see how they map onto a single
fixed value of $w$.  An obvious first choice are models described by the CPL parametrization (Eq. \ref{CPL}).
Beyond that case, there are an infinite set of models to choose from, so we have limited our investigation
to the thawing hilltop quintessence models discussed in Ref. \cite{ds1} (see also the extensions to
this approach given in Refs. \cite{Chiba,ds2}).
These models are characterized by a potential that is well approximated as an inverted harmonic oscillator:
\begin{equation}
V(\phi) = V(\phi_m) + (1/2) V^{\prime \prime}(\phi_m) \phi^2,
\end{equation}
where $\phi_m$ is the value of $\phi$ at which the potential achieves its maximum.
Ref. \cite{ds1} showed that in a background expansion close to $\Lambda$CDM, scalar field models for which the field $\phi$
is evolving near the maximum of the potential tend to converge toward a similar evolution with the scale factor, namely the
equation of state parameter is well-approximated by
\begin{equation}
\label{finalfinal}
1 + w(a) = (1+w_0)a^{3(K-1)}\frac{[(F(a)+1)^K(K-F(a))
+(F(a)-1)^K(K+F(a))]^2}
{[(\Omega_{\phi0}^{-1/2}+1)^K(K-\Omega_{\phi0}^{-1/2})
+(\Omega_{\phi0}^{-1/2}-1)^K (K+\Omega_{\phi0}^{-1/2})]^{2}},
\end{equation}
where $F(a)$ is given by
\begin{equation}
F(a) = \sqrt{1+(\Omega_{\phi 0}^{-1}-1)a^{-3}},
\end{equation}
and the constant $K$ characterizes the curvature of the potential near its maximum:
\begin{equation}
\label{Kdef}
K = \sqrt{1-(4/3)V^{\prime \prime}(\phi_m)/V(\phi_m)}.
\end{equation}
While the general form for $w$ as a function of $a$ is rather complex, it simplifies
considerably for integer values of $K$, reducing to \cite{ds1}
\begin{eqnarray}
\label{K2}
K&=&2:~~1+w = (1+w_0) a^3,\\
\label{K3}
K&=&3:~~1+w = (1+w_0)[(1-\Omega_{\phi0})a^3 + \Omega_{\phi0}a^6],\\
\label{K4}
K&=&4:~~1+w = \frac{1+w_0}{(5+\Omega_{\phi0})^2}\left[
25(1-\Omega_{\phi0})^2 a^3 + 60\Omega_{\phi0}(1-\Omega_{\phi 0})a^6
+ 36 \Omega_{\phi 0}^2 a^9 \right].
\end{eqnarray}
While Shlivko and Steinhardt {\cite{Shlivko} also examined thawing models
of this type, they elected to use an exact numerical calculation of the evolution.
This obviously provides a more accurate treatement for $w(a)$.
However, our goal here is not to claim that any of these models provides the true description
for the evolution of dark energy.  Rather, we wish to examine a set of analytically tractable models
that diverge from both the CPL parametrization and constant $w$ evolution to see if there
is any pattern to the way that they map to constant $w$.  The models discussed here are ideal
for this purpose.                                                                                                                            
                                                                                                                                  
Consider first the models described by the CPL parametrization.	From Eqs. (\ref{CPL}) and (\ref{rho(z)})
we derive the standard expression for the evolution of the density of dark energy described by the CPL equation
of state parameter:
\begin{equation}
\rho(a)/\rho_{\phi 0} = a^{-3(1+w_0+w_a)}e^{3w_a(a-1)}.
\end{equation}
Using the procedure outlined above, we determine the best-fit $w_*$ corresponding to each pair of values $(w_0, w_a)$.
This mapping is displayed in Fig. 1.  While this mapping appears too complex to provide any useful information,
this apparent complexity hides a very simple result.  If we calculate the redshift $z_{pivot}$ corresponding to
$w_*$ in the CPL parametrization, i.e. $w_* = w_0 + w_a[1-1/(1+z_{pivot})]$, then we see that $z_{pivot}$ is nearly
independent of $w_0$ and $w_a$, as shown in Fig. 2.  Thus, fitting dark energy that evolves exactly as
given by the CPL parametrization to a fixed value of $w$ picks out the value of $w$ achieved by the dark
energy at the redshift $z_{pivot} = 0.22 - 0.25$.

This result is not surprising.  It has long been known that supernovae data are most sensitive to the
value of the equation of
state parameter at $z \approx 0.2$ \cite{Huterer,Maor2}. This is closely related to the existence
of a ``pivot redshift" when fitting the CPL parametrization to supernovae data; at the pivot redshift
the best-fit values of $w_0$ and $w_a$ are uncorrelated and the uncertainty in $w$ is minimized
\cite{Huterer,HuJain,Albrecht,Linderpivot,Martin}.  Note, however, that our results for the CPL parametrization do not
map to a single value of $z$, but a narrow range of values, so we will refer to this as a ``pivot-like" redshift.
	
\begin{figure}[!ht]
\centering
\includegraphics[width= 5in]{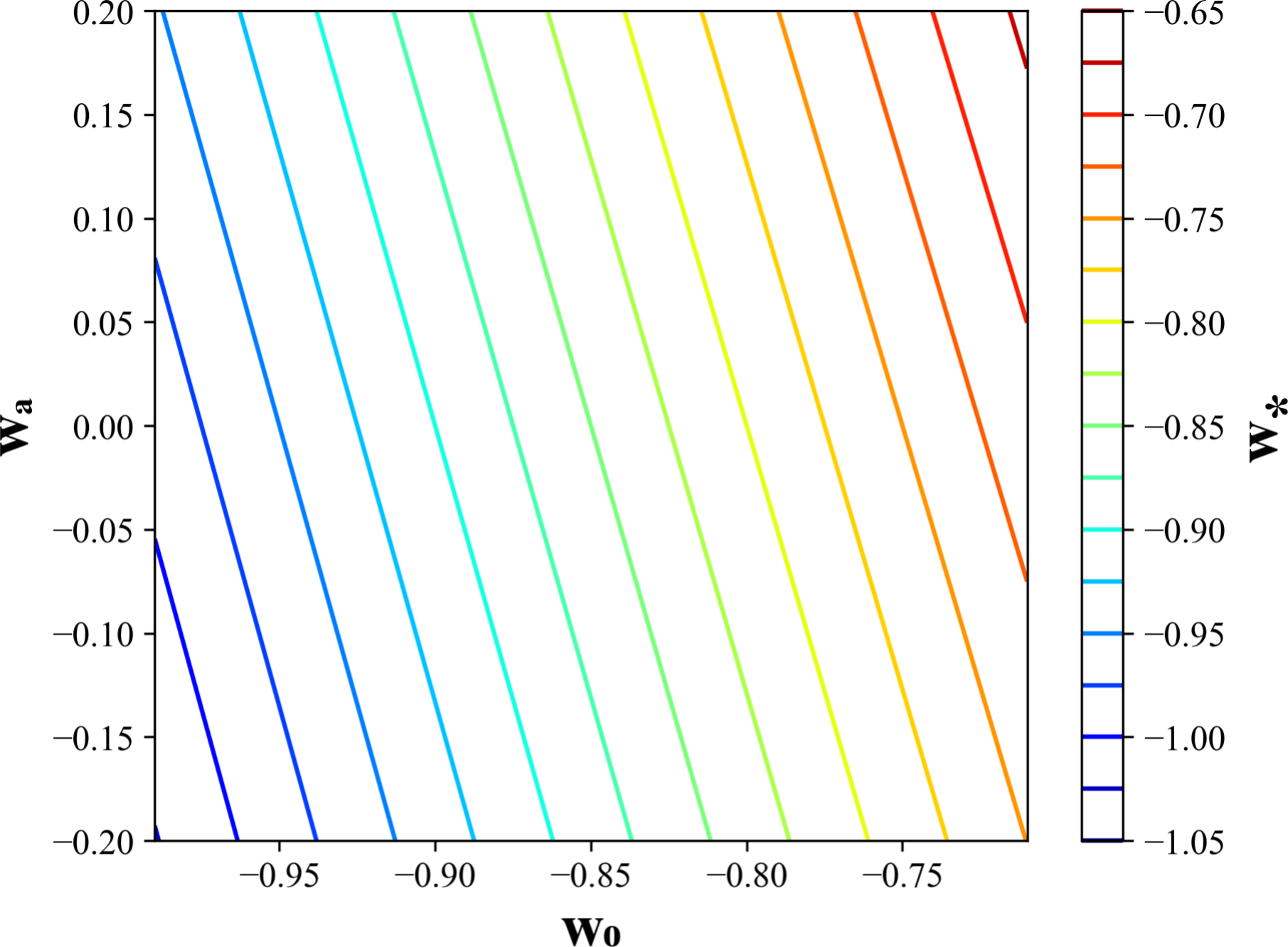}
\caption{The value of $w_*$ for a dark energy model with a constant equation of state parameter that gives
the best fit to the distance modulus produced by a true dark energy model with evolving equation of
state parameter described by the CPL parametrization,
$w = w_0 + (1-a) w_a$, for the indicated values of $w_0$ and $w_a$.}
\end{figure}	
	
\begin{figure}[!ht]
\centering
\includegraphics[width= 5in]{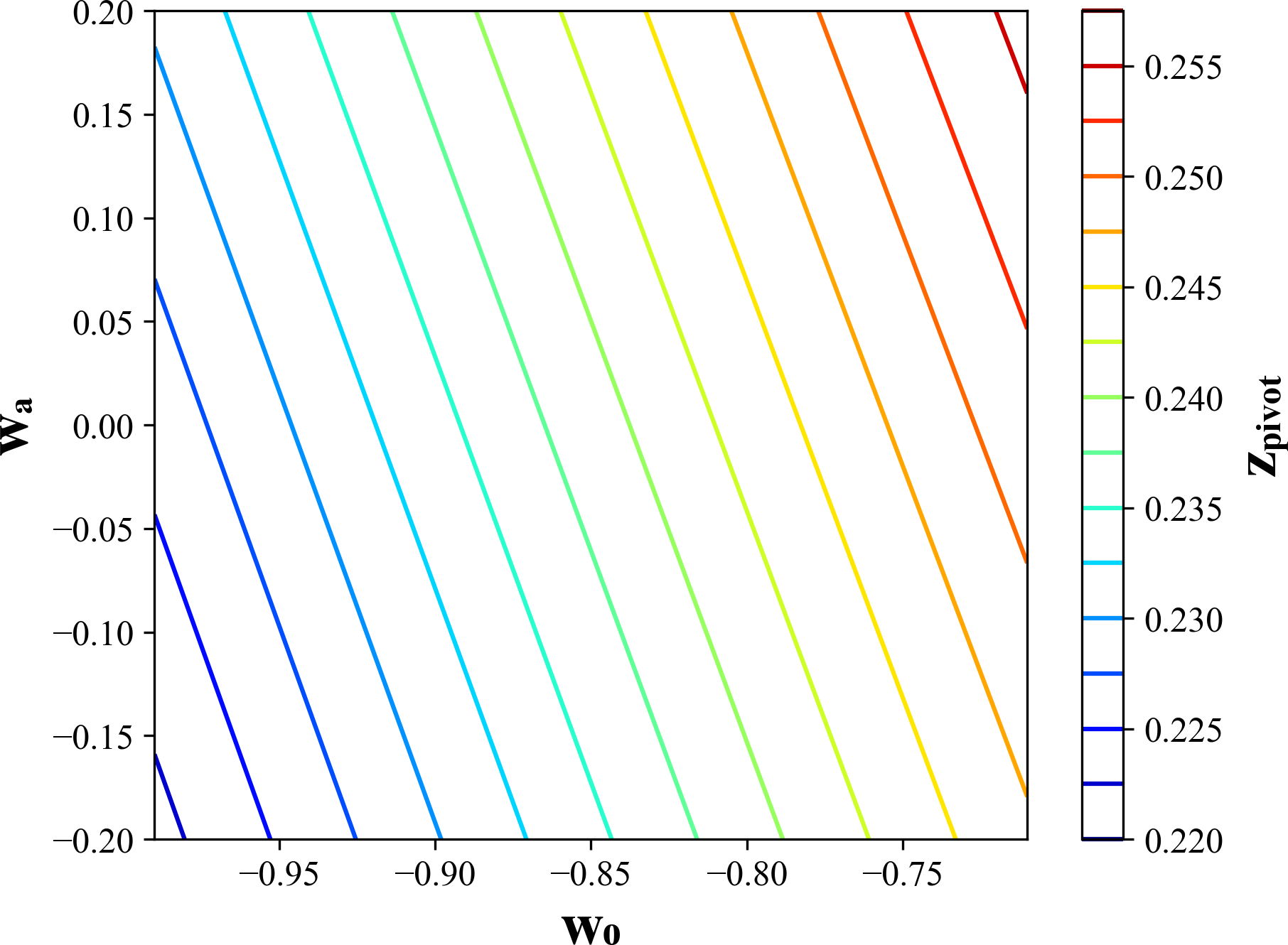}
\caption{The pivot-like redshift, $z_{pivot}$, at which the value of $w$ in the CPL parametrization is equal to the
best-fit $w_*$ in Fig. 1.}
\end{figure}			

\begin{figure}[!ht]
\centering
\includegraphics[width= 5in]{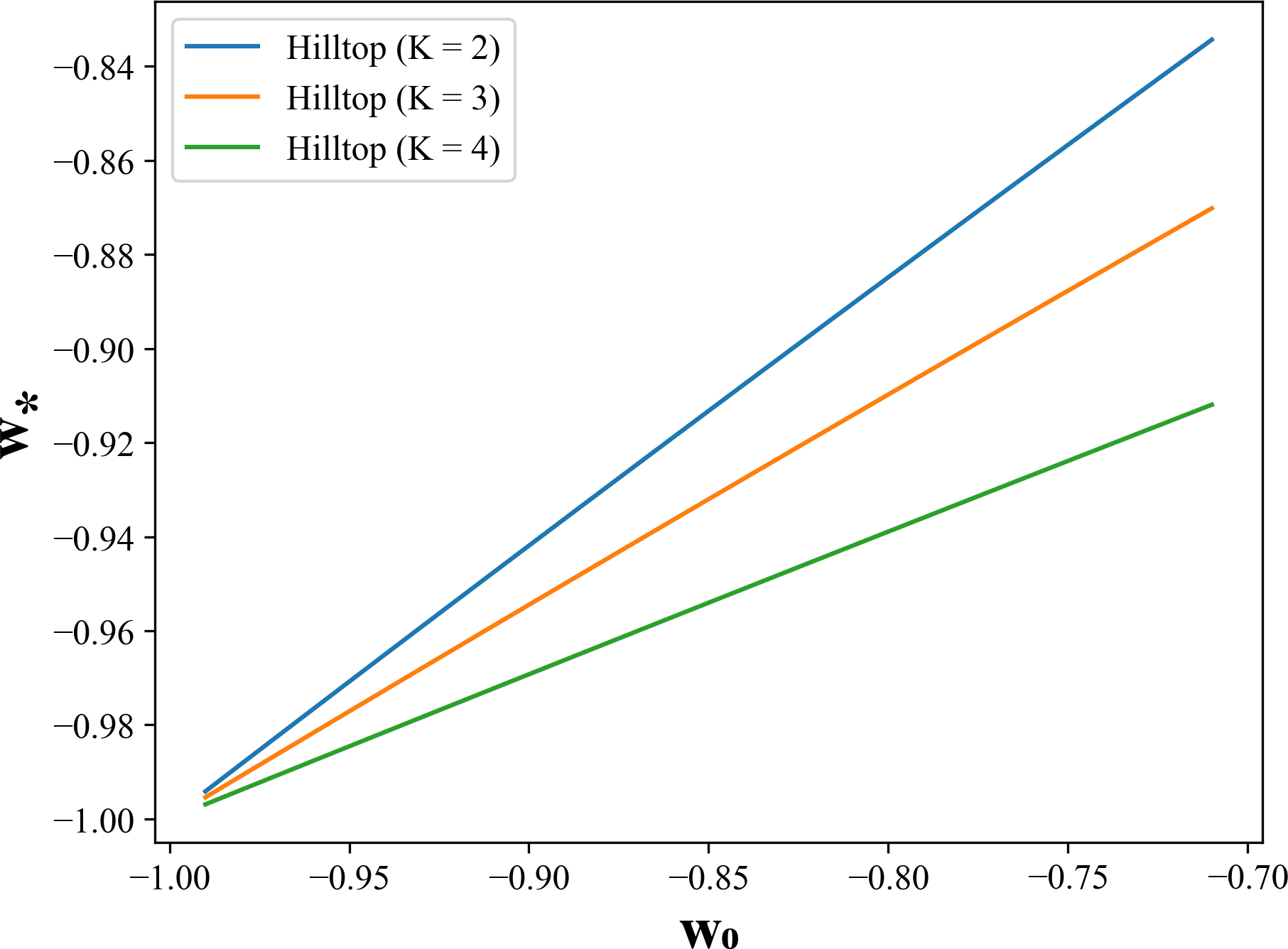}
\caption{The value of $w_*$ for a dark energy model with a constant equation of state parameter that gives
the best fit to the distance modulus produced by a true hilltop quintessence model with evolving equation of state parameter
for the indicated values of $w_0$ and $K$ in Eqs. (\ref{K2})$-$(\ref{K4}).}.
\end{figure}	

\begin{figure}[!ht]
\centering
\includegraphics[width= 5in]{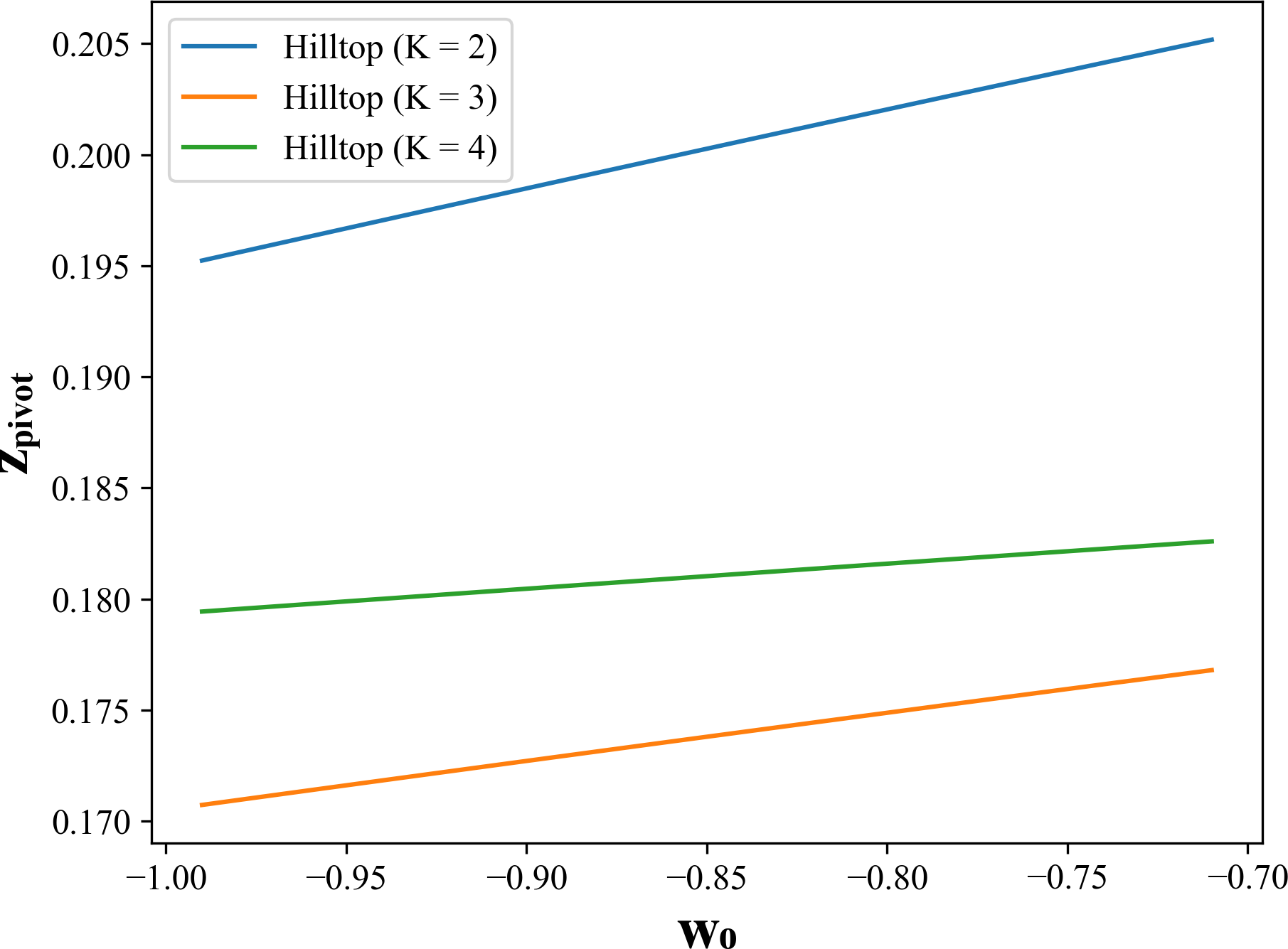}
\caption{The pivot-like redshift,
$z_{pivot}$, at which the value of $w$ given by the hilltop quintessence model with the indicated
values of $w_0$ and $K$ is equal to the best-fit value of
$w_*$ in Fig. 3.}
\end{figure}	

Now consider the more complex thawing quintessence models given by
Eqs. (\ref{K2})$-$(\ref{K4}).  These models all evolve
from an initial value of $w \approx -1$ to a final value
of $w = w_0$, with higher curvature in the potential (larger $K$) giving later and more
rapid evolution to the present-day value of $w$. (See Fig. 1 of Ref. \cite{ds1}). Note that the value $K=1$ corresponds to a thawing quintessence model
in a nearly linear potential.  Quintessence models with a linear potential were explored in detail in Ref. \cite{ScherrerSen},
where it was shown that they are well-approximated by the CPL parametrization with $w_a = -1.5(1+w_0)$,
so the $K=1$ models represent a subset of the already-examined CPL models.

For larger values of $K$, we use Eqs. (\ref{K2})$-$(\ref{K4})
to derive the corresponding expressions for $\rho_\phi(a)/\rho_{\phi0}$:
\begin{eqnarray}
\label{K2rho}
K&=&2:~~\rho_\phi(a)/\rho_{\phi0} = \exp[(1+w_0)(1-a^3)],\\
\label{K3rho}
K&=&3:~~\rho_\phi(a)/\rho_{\phi0} =
\exp\left\{(1+w_0)\left[(1-\Omega_{\phi0})(1-a^3) + \frac{1}{2}
\Omega_{\phi0}(1-a^6)\right]\right\},\\
\label{K4rho}
K&=&4:~~\rho_\phi(a)/\rho_{\phi0} = \exp\left\{\frac{1+w_0}{(5+\Omega_{\phi0})^2}\left[
25(1-\Omega_{\phi0})^2 (1-a^3) + 30\Omega_{\phi
0}(1-\Omega_{\phi 0})(1-a^6)
+ 12 \Omega_{\phi 0}^2 (1-a^9) \right]\right\}.
\end{eqnarray}
Using these values for $\rho_\phi/\rho_{\phi 0}$, we derive $w_*$, the constant value of $w$
that gives the best fit to the distance modulus for the hilltop models.  The values of $w_*$ as a function of $w_0$
for  $K = 2-4$ are shown
in Fig. 3.

We see that $1+w_*$ is almost exactly linearly proportional to $1+w_0$, with a slope that decreases with
increasing $K$.  This is easy to understand on physical grounds.  For a fixed
value of $w_0$, a larger value of $K$ corresponds to a $w(a)$ that evolves later and more rapidly
from $w = -1$ to $w = w_0$.  Thus, larger $K$ at fixed $w_0$ corresponds to a value of $w(a)$ that is smaller
over the entire range of $a$ (see Fig. 1 of Ref. \cite{ds1}), which must yield a smaller value of $w_*$.

The fact that $1+w_*$ is nearly linearly proportional to $1+w_0$, combined with
Eqs. (\ref{K2}) $-$ (\ref{K4}), indicates that
these approximations to the hilltop models yield a pivot-like value of the redshift, $z_{pivot}$
at which the value of $1+w$ given by Eqs. (\ref{K2}) $-$ (\ref{K4}) is equal to $1+w_*$, where
$z_{pivot}$ depends on $K$ but is nearly independent of $w_0$.  The value of
$z_{pivot}$ is shown in Fig. 4.  Note that $z_{pivot}$ does not
vary monotically with $K$; the pivot-like redshift for $K=4$ lies between the values for $K=2$ and $K=3$.  
Further, $z_{pivot}$ is not exactly constant for a given $K$ but instead varies slowly with $w_0$.

\section{Discussion}

For all of the models examined here with time-varying equations of state, a fit to a constant
equation of state parameter simply
provides the value of $w$ at a pivot-like redshift in the range $0.17 - 0.25$, with the exact value of $z_{pivot}$
depending on the particular model.  
For models in which $w$ varies linearly with the scale factor
(CPL models)
this result was already suggested by earlier work; we have extended it here
to a more general class of models with nonlinear evolution of $w$.
For the latter models, the behavior of $z_{pivot}$ as a function of the model parameters
is not straightforward.  We find that $z_{pivot}$ for the hilltop models varies only very slowly 
with $w_0$, the present-day value of $w$, but it is quite sensitive to the hilltop curvature in these models
as parametrized by $K$.  Furthermore, $z_{pivot}$ is not a monotonic function of the curvature of the scalar
field potential.  

These results extend the earlier work by Maor et al. \cite{Maor2}, who showed the limitations of assuming a fixed value
of $w$.
The obvious question is whether any useful information is provided by fitting supernovae data to a constant value
of $w$.  Our results are somewhat ambiguous in that regard.  It is clear that for the range of models examined
here, fitting a model with a time-varying equation of state parameter to a fixed value of $w$ simply picks
out the value of $w$ at a redshift near $z = 0.2$.  The fact that this is the case for the full range of models
considered here is somewhat useful.  However, the precise value of $z_{pivot}$ varies over the models
we have examined, ranging from $0.17$ to $0.20$ for the hilltop thawing models and from $0.22$ to $0.25$
for the CPL models.  Hence, we must conclude that a constant $w$ fit provides only limited information.
While
we have examined only a particular set of models with time-varying $w$, we already see major variation in
$z_{pivot}$ within these models, so an extension of this study to a larger set of models seems unwarranted.

It is important to note the limitations of this study.  In order to focus exclusively on the behavior of fits to supernovae
data, we have neglected other data sets that are normally used to measure $w$, such as the cosmic microwave
background (CMB) and baryon acoustic oscillations (BAO).  The inclusion of these other data sets would necessarily
modify the best-fit value of $w$, as highlighted recently by Perez et al. \cite{PPR}.  For similar reasons,
we have fixed the value of $\Omega_\phi$ and taken the curvature to be zero, rather than treating them
as free parameters.
By fixing $\Omega_\phi$ in these models we have enhanced
their predictive power; allowing $\Omega_\phi$ to vary
would significantly reduce the information we have derived from the constant-$w$ fits, as noted
in Ref. \cite{Maor2}.  Similarly, the inclusion of curvature would alter the best-fit value of $w$, as can be seen
in the results of Ref. \cite{PPR}.

The data that support the findings of this article are openly available \cite{repository}.

\end{document}